\documentclass[a4paper,11pt]{article}
\usepackage{pos}
\renewcommand{\logo}{\relax}

\numberwithin{equation}{section}
\usepackage{physics2}
\usephysicsmodule{ab, ab.braket}
\newcommand{\order}[1]{\mathcal{O} \ab( #1 )}
\newcommand{\eval}[1]{\left. #1 \right|}
\usepackage{fixdif}
\usepackage{derivative}
\NewDifferential{\dd}{\mathrm{d}}[sep-end=\;]
\usepackage{slashed}
\newcommand{\nbar}{\bar{n}}
\newcommand{\gt}{g_T}
\newcommand{\psibar}{\bar{\psi}}

\title{Tensor-polarized twist-3 distribution function $f_{LT}$ of spin-1 deuteron}

\author[a,b,c]{S.~Kumano}
\author*[a,b]{Kenshi~Kuroki}

\affiliation[a]{Quark Matter Research Center, Institute of Modern Physics, Chinese Academy of Sciences,\\
509 Nanchang Rd., Lanzhou, 730000, China}

\affiliation[b]{Southern Center for Nuclear Science Theory, Institute of Modern Physics, Chinese Academy of Sciences,\\
No.~1 Xinqiao North Rd., Huizhou, 516000, China}

\affiliation[c]{KEK Theory Center, Institute of Particle and Nuclear Studies, KEK,\\
Oho 1-1, Tsukuba, 305-0801, Japan}

\emailAdd{kumanos@impcas.ac.cn}
\emailAdd{k-kuroki@impcas.ac.cn}

\abstract{
We investigate the twist-2 and twist-3 tensor-polarized partonic structures of the spin-1 deuteron.
Using the operator product expansion with local operators, we derive a Wandzura-Wilczek (WW)-like twist-2 relation between the tensor-polarized twist-2 quark distribution $f_{1LL}$ and the twist-3 distribution $f_{LT}$, together with a Burkhardt-Cottingham (BC)-like sum rule.
The local-operator formalism makes the Lorentz structure and rotational invariance manifest and provides useful constraints on the twist-3 sector.
We then use a phenomenological parametrization of the twist-2 distribution $f_{1LL}(x)$, constrained by HERMES data on the deuteron structure function $b_1$ at $Q^2=2.5~\text{GeV}^2$, to quantitatively estimate the twist-3 distribution $f_{LT}(x)$ within the WW approximation.
The resulting $f_{LT}(x)$ has a shape and magnitude comparable to those of $f_{1LL}(x)$, indicating that subleading-twist effects may be relevant for future experiments at relatively low $Q^2$.
In this contribution, we summarize the theoretical derivation of the WW-like relation and BC-like sum rule and present a phenomenological estimate of the tensor-polarized twist-3 quark distribution $f_{LT}$ in the deuteron.
}

\FullConference{The 33rd International Workshop on Deep Inelastic Scattering and Related Subjects (DIS2026)\\
4 - 8 May 2026\\
Bologna, Italy\\}

\begin{document}
\maketitle

\section{Introduction}

Deep-inelastic scattering (DIS) of spin-1/2 nucleons has played a central role in establishing the partonic description of hadrons.
For a spin-1 target, however, the spin density matrix contains additional degrees of freedom associated with tensor polarization.
Consequently, spin-1 hadrons possess parton distribution functions (PDFs) and structure functions with no direct analogues for spin-1/2 targets.
The deuteron is a stable spin-1 hadronic system and therefore provides a particularly useful target for studying this additional spin structure.
In charged-lepton inclusive DIS on tensor-polarized spin-1 targets, there are four tensor-polarized structure functions~\cite{Hoodbhoy:1988am}: $b_1, b_2, b_3,$ and $b_4$. 
The structure function $b_1$ was measured for the deuteron by the HERMES Collaboration in 2005~\cite{HERMES:2005pon}, providing the first experimental information on tensor-polarized partonic structure.
A tensor-polarized deuteron program at the Thomas Jefferson National Accelerator Facility (JLab) is also being prepared, while measurements at future Electron-Ion Colliders and hadron facilities may provide complementary access to spin-1 structure.

The structure function $b_1$ is particularly interesting because its measured behavior is not well reproduced by conventional deuteron models based on a nucleonic convolution picture~\cite{Khan:1991qk, Cosyn:2017fbo, Kumano:2026ipc}, a phenomenon often referred to as the ``tensor-polarization puzzle''.
This motivates a more detailed investigation of tensor-polarized PDFs, particularly their higher-twist contributions.
For a spin-1 hadron, the collinear quark correlation function contains three tensor-polarized distributions up to twist 3~\cite{Kumano:2020ijt}: $f_{1LL}, e_{LL}$, and $f_{LT} $.
Here, $f_{1LL}$ is twist 2, and $e_{LL}$ and $f_{LT}$ are twist 3.
Among these distributions, $e_{LL}$ is chiral odd and therefore does not contribute to inclusive DIS.
The remaining distributions, the twist-2 $f_{1LL}$ and twist-3 $f_{LT}$, are related by a Wandzura-Wilczek (WW)-like twist-2 relation and satisfy a Burkhardt-Cottingham (BC)-like sum rule~\cite{Kumano:2021fem}.

In the following, we first briefly review the derivation of the twist-2 relation and the associated sum rule using the operator product expansion (OPE) method with local operators, which makes the Lorentz structure and rotational invariance explicit.
Second, we use the resulting relation together with a phenomenological parametrization of the twist-2 distribution constrained by the HERMES $b_1$ data~\cite{Kumano:2010vz} to obtain a quantitative estimate of the twist-3 distribution $f_{LT}$ in the deuteron.
The main results presented here are based on Refs.~\cite{Kumano:2026xxv, Kumano:2025rai}.

\section{Twist-2 relation and sum rule for tensor-polarized distribution functions}

We first introduce the tensor-polarized PDFs and derive the relation between $f_{1LL}$ and $f_{LT}$.
For details, see Ref.~\cite{Kumano:2026xxv}.
Throughout this contribution, we consider only the tensor-polarized parts and neglect the unpolarized and vector-polarized parts, which are essentially the same as those for spin-1/2 nucleons.
The PDFs are defined through the quark collinear correlation function involving a non-local operator
\begin{equation}
    \Phi_{ij} \ab(x_D;P,T) = \eval{ \int \frac{\d{\ab(P\cdot\xi)}}{2\pi P^+} e^{ix_DP\cdot\xi} \braket<P,T|\psibar_j(0)\psi_i(\xi)|P,T> }_{\xi^+=0,\, \vec{\xi_T}=0},
\end{equation}
where $\psi$ denotes the quark field, while $P$ and $T$ denote the momentum and spin tensor of the deuteron, respectively.
The lightcone $\pm$ components are defined as $a^\pm = \ab(a^0 \pm a^3)/\sqrt{2}$, and $x_D$ denotes the fraction of the deuteron momentum carried by the quark and is defined in the range $0 \leq x_D \leq 1$.
The gauge link required for color-gauge invariance is not shown explicitly.
By introducing lightcone vectors $n^\mu=\ab(1,0,0,-1)/\sqrt{2}$, $\nbar^\mu=\ab(1,0,0,1)/\sqrt{2}$, and transverse metric $\gt^{\mu\nu}=g^{\mu\nu}-n^{\mu}\nbar^{\nu}-\nbar^\mu n^\nu$, the spin tensor is expressed using parameters $S_{LL}, S_{LT}^\mu$, and $S_{TT}^{\mu\nu}$ as~\cite{Bacchetta:2000jk}
\begin{equation}
  \begin{split}
    T^{\mu\nu} =&\, \frac{2\ab(P^+)^2}{3M_D^2} S_{LL} \nbar^\mu \nbar^\nu
               - \frac{1}{3} S_{LL} \ab( \nbar^\mu n^\nu + \nbar^\nu n^\mu -\gt^{\mu\nu} )
               + \frac{M_D^2}{6\ab(P^+)^2} S_{LL} n^\mu n^\nu \\
               &\quad + \frac{P^+}{2M_D} \ab( \nbar^\mu S_{LT}^\nu + \nbar^\nu S_{LT}^\mu )
               - \frac{M_D}{4P^+} \ab( n^\mu S_{LT}^\nu + n^\nu S_{LT}^\mu )
               + S_{TT}^{\mu\nu},
  \end{split}
\end{equation}
where $M_D$ is the deuteron mass.
The parameters satisfy the relations: $S_{LT}\cdot\nbar = S_{LT}\cdot n = 0$, $S_{TT}^{\mu\nu}\nbar_\nu = S_{TT}^{\mu\nu}n_\nu = 0$, and $S_{TT}^{\mu\nu}g_{T\mu\nu} = 0$.
By using these parameters, the correlation function is expressed in terms of collinear PDFs as~\cite{Kumano:2020ijt}
\begin{equation}
    \Phi\ab(x_D,P,T) = \frac{1}{2} S_{LL} \slashed{\nbar} f_{1LL}(x_D)
                     + \frac{M_D}{2P^+} S_{LL} e_{LL}(x_D)
                     + \frac{M_D}{2P^+} \slashed{S}_{LT} f_{LT}(x_D)
                     + \order{\frac{M_D^2}{\ab(P^+)^2}},
\end{equation}
where $\slashed{a} = \gamma^\mu a_\mu$ with Dirac matrices $\gamma^\mu$.

To find the twist-2 relation, we first consider the $n$-th moment of the correlation function of the non-local vector operator $\psibar(0)\gamma^\sigma\psi(\xi)$, which relates to the moments of PDFs $f_{1LL}$ and $f_{LT}$ as
\begin{equation}\label{eq:CF_moment}
  \begin{split}
    &\int_{-1}^{1}\dd{x_D} x_D^{n-1} \eval{ \int \frac{\d\ab(P\cdot\xi)}{4\pi P^+} e^{ix_DP\cdot\xi} \braket<P,T|\psibar(0)\gamma^\sigma\psi(\xi)|P,T> }_{\xi^+=0,\, \vec{\xi_T}=0} \\
    &= S_{LL} \nbar^\sigma \int_{-1}^{1}\dd{x_D} x_D^{n-1} f_{1LL}(x_D)
    + \frac{M_D}{P^+} S_{LT}^\sigma \int_{-1}^{1}\dd{x_D} x_D^{n-1} f_{LT}(x_D).
  \end{split}
\end{equation}
While the scaling variable $x_D$ is restricted to the physical region $0\leq x_D \leq 1$ for quark distributions, the moments are taken over $-1 \leq x_D \leq 1$ to include both quark and antiquark contributions.
Expanding the quark field $\psi(\xi)$ around $\xi=0$ and using the relation
\begin{equation}
    \partial^{\mu_1} \cdots \partial^{\mu_{n-1}} \sum_{m=0}^{\infty} \frac{1}{m!} \ab(\xi\cdot\partial)^m
    = \partial^{\{\mu_1} \cdots \partial^{\mu_{n-1}\}} \ab( 1 + \sum_{m=1}^{\infty} \frac{1}{m!} \ab(\xi\cdot\partial)^m ),
\end{equation}
where the curly brackets denote complete symmetrization of the Lorentz indices defined by $ \partial^{\{\mu_1} \cdots \partial^{\mu_{n-1}\}} = \ab( \partial^{\mu_1} \cdots \partial^{\mu_{n-1}} + \text{permutations} ) / \ab(n-1)!$, the left-hand side of Eq.~\eqref{eq:CF_moment} can also be written in terms of the local operator as~\cite{Kumano:2026xxv}
\begin{gather}
    \begin{split}
        &\int_{-1}^{1}\dd{x_D} x_D^{n-1} \eval{ \int \frac{\d\ab(P\cdot\xi)}{4\pi P^+} e^{ix_DP\cdot\xi} \braket<P,T|\psibar(0)\gamma^\sigma\psi(\xi)|P,T> }_{\xi^+=0,\, \vec{\xi_T}=0} \\
        &= \frac{n_{\mu_1}\cdots n_{\mu_{n-1}}}{2\ab(P^+)^n} \braket<P,T|R^{\sigma\{\mu_1\cdots\mu_{n-1}\}}|P,T>,
    \end{split} \label{eq:ME_local} \\[0.5cm]
    R^{\sigma\{\mu_1\cdots\mu_{n-1}\}} = i^{n-1} \psibar(0) \gamma^\sigma D^{\{\mu_1} \cdots D^{\mu_{n-1}\}} \psi(0) - \text{traces},
\end{gather}
where $D$ denotes the covariant derivative.
The local operator can be decomposed into twist-2 completely symmetric and twist-3 mixed symmetric operators as $R^{\sigma\{\mu_1\cdots\mu_{n-1}\}} = R^{\{\sigma\mu_1\cdots\mu_{n-1}\}} + R^{[\sigma\{\mu_1]\cdots\mu_{n-1}\}}$, where the square brackets $[\cdots]$ denote antisymmetrization of indices.
By expressing the matrix elements of the local operators using momentum $P^\mu$ and spin tensor $T^{\mu\nu}$ as
\begin{align}
    \begin{split}
        \braket<P,T|R^{\{\sigma\mu_1\cdots\mu_{n-1}\}}|P,T>
        &= \frac{2}{n} a_n M_D^2 \left[ \sum_{i=1}^{n-1} \ab(T^{\sigma\mu_i}+T^{\mu_i\sigma}) \prod_{j(\neq i)=1}^{n-1} P^{\mu_j} \right. \\
        &\hspace{2.5cm} \left. + \sum_{i=1}^{n-1} \sum_{j(\neq i)=1}^{n-1} T^{\mu_i\mu_j}P^\sigma \prod_{k(\neq i,j)=1}^{n-1} P^{\mu_k} \right], 
    \end{split} \\[0.3cm]
    \braket<P,T|R^{[\sigma\{\mu_1]\cdots\mu_{n-1}\}}|P,T>
    &= \frac{2}{n} d_n M_D^2 \sum_{i=1}^{n-1} \sum_{j(\neq i)=1}^{n-1} \ab(P^\sigma T^{\mu_i\mu_j} - P^\mu T^{\sigma\mu_j}) \prod_{k(\neq i,j)=1}^{n-1} P^{\mu_k},
\end{align}
where the overall coefficients $a_n$ and $d_n$ correspond to twist-2 and twist-3 contributions, respectively, the right-hand side of Eq.~\eqref{eq:ME_local} becomes
\begin{equation}\label{eq:ME_local_2}
  \begin{split}
    &\frac{n_{\mu_1}\cdots n_{\mu_{n-1}}}{2\ab(P^+)^n} \braket<P,T|R^{\sigma\{\mu_1\cdots\mu_{n-1}\}}|P,T> \\
    &= S_{LL} \nbar^\sigma \ab[\frac{2(n-1)}{3} a_n]
    + \frac{M_D}{P^+} S_{LT}^\sigma \ab[\frac{n-1}{n} a_n - \frac{(n-1)(n-2)}{2n} d_n].
  \end{split}
\end{equation}
By comparing the coefficients of the independent tensor structures on the right-hand side of Eqs.~\eqref{eq:CF_moment} and \eqref{eq:ME_local_2}, we obtain the moments of the PDFs
\begin{alignat}{4}
    &\int_{-1}^1 \dd{x_D}& x_D^{n-1} &f_{1LL}(x_D) &&= \,&\frac{2(n-1)}{3}&\, a_n, \label{eq:f1LL_moment} \\
    &\int_{-1}^1 \dd{x_D}& x_D^{n-1} &f_{LT}(x_D)  &&= \,&\frac{n-1}{n}&\, a_n \,-\,\frac{(n-1)(n-2)}{2n} d_n. \label{eq:fLT_moment}
\end{alignat}
The first term on the right-hand side of Eq.~\eqref{eq:fLT_moment} originates from the twist-2 operator, whereas the second term represents the twist-3 operator associated with multiparton correlations.
Thus, the twist-3 distribution $f_{LT}$ can be decomposed into a WW-like twist-2 part $f_{LT}^\text{twist-2}$ and a genuine twist-3 part $f_{LT}^\text{(HT)}$ as
\begin{gather}
    f_{LT}(x_D) = f_{LT}^\text{twist-2}(x_D) + f_{LT}^\text{(HT)}(x_D), \\[0.3cm]
    \int_{-1}^1 \dd{x_D} x_D^{n-1} f_{LT}^\text{twist-2}(x_D) = \frac{n-1}{n} a_n, \label{eq:fLTWW_moment} \\
    \int_{-1}^1 \dd{x_D} x_D^{n-1} f_{LT}^\text{(HT)}(x_D) = - \frac{(n-1)(n-2)}{2n} d_n.
\end{gather}

Finally, Eqs.~\eqref{eq:f1LL_moment} and \eqref{eq:fLTWW_moment} lead to the following WW-like twist-2 relation for the tensor-polarized PDFs:
\begin{equation}\label{eq:WW-like}
    f_{LT}^\text{twist-2}(x_D) = \frac{3}{2} \int_{x_D}^1 \frac{\d y_D}{y_D} f_{1LL}(y_D),
\end{equation}
where the WW-like part $f_{LT}^\text{twist-2}$ is completely determined by the twist-2 distribution $f_{1LL}$\@.
One also finds the BC-like sum rule:
\begin{equation}\label{eq:BC-like}
    \int_0^1 \dd{x_D} f_{LT}^\text{twist-2}(x_D) - \frac{3}{2} \int_0^1 \dd{x_D} f_{1LL}(x_D) = 0.
\end{equation}

\section{Twist-3 distribution function \texorpdfstring{$f_{LT}$}{fLT} of the deuteron}

Using the WW-like twist-2 relation in Eq.~\eqref{eq:WW-like}, the twist-3 PDF $f_{LT}$ can be obtained from the twist-2 PDF $f_{1LL}$ by neglecting the genuine twist-3 contribution $f_{LT}^\text{(HT)}$.
This approximation is commonly referred to as the Wandzura-Wilczek approximation.
In this section, we quantitatively estimate the twist-3 PDF $f_{LT}$ of the deuteron.
The twist-3 distribution $f_{LT}$ can in principle be probed in various processes, including inclusive DIS, semi-inclusive DIS (SIDIS)~\cite{Zhao:2025vol}, and Drell-Yan processes~\cite{Qiao:2024bgg}.

In Ref.~\cite{Kumano:2010vz}, the leading-twist tensor-polarized distributions were determined from a $\chi^2$ fit to HERMES data for the deuteron structure function $b_1$ at $Q^2 = 2.5 \text{GeV}^2$.
For the deuteron, we use the scaling variable $x$ defined with the nucleon mass, for which the kinematic range is $0 \leq x \leq 2$.
The resulting valence- and sea-quark distributions $f_{1LL}$ are parametrized as
\begin{align}
    f_{1LL}^{q_v/D}(x)     &= -\frac{2}{3}                  \delta_T w(x) q_v^D(x), \\
    f_{1LL}^{\bar{q}/D}(x) &= -\frac{2}{3} \alpha_{\bar{q}} \delta_T w(x) \bar{q}^D(x), \\[0.3cm]
    \delta_T w(x)          &= a x^b \ab(1-x)^c \ab(x_0 - x),
\end{align}
with parameter values $a=0.221\pm0.174, b=0.648\pm0.342, \alpha_{\bar{q}}=3.20\pm2.75, c=1 (\text{fixed})$, and $x_0=0.221$.
Here, $q_v^D$ and $\bar{q}^D$ denote the unpolarized PDFs of the deuteron per nucleon, which are constructed from the MSTW 2008 LO proton PDFs~\cite{Martin:2009iq} according to $q_v^D=\ab(u_v+d_v)/2$ and $\bar{q}^D=\ab(2\bar{u}+2\bar{d}+s+\bar{s})/6$.
\begin{figure}[tbp]
  \centering
  \begin{minipage}[t]{0.49\hsize}
    \includegraphics[clip,width=\linewidth]{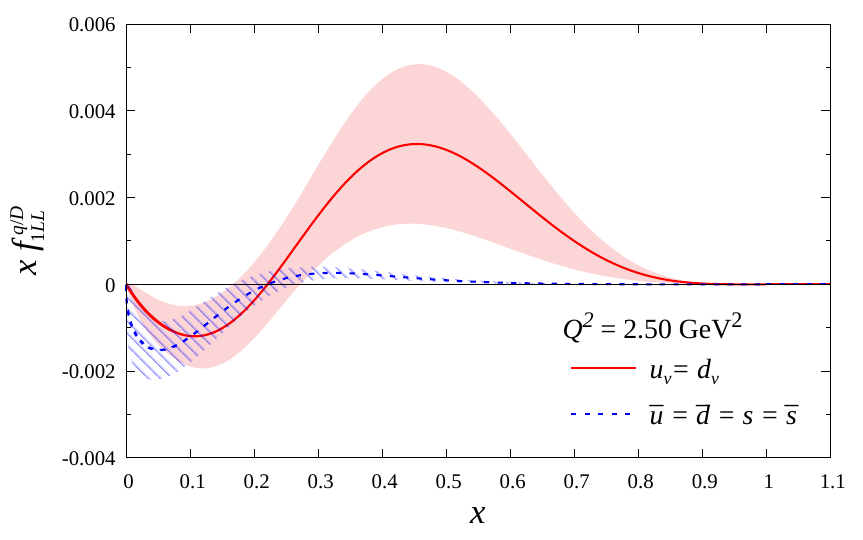}
  \end{minipage}
  \begin{minipage}[t]{0.49\hsize}
    \includegraphics[clip,width=\linewidth]{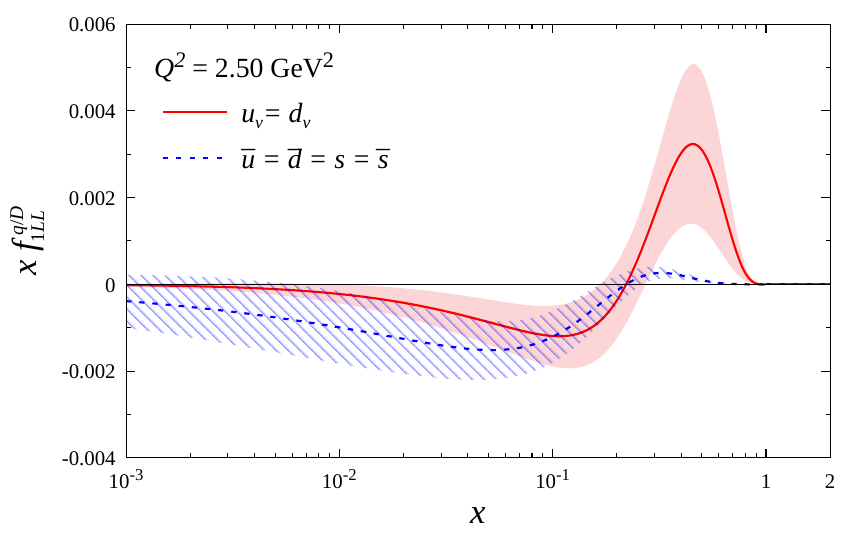}
  \end{minipage}
  \caption{
    Tensor-polarized twist-2 distributions $f_{1LL}$ of the deuteron.
    The red solid lines represent the valence-quark distributions, while the blue dashed lines represent the sea-quark distributions.
    The uncertainty bands correspond to $\varDelta\chi^2=1$.
    The left panel uses a linear scale for Bjorken-$x$, while the right panel uses a semi-logarithmic scale to better display the small-$x$ region.
  }\label{fig:f1LL}
\end{figure}
Figure~\ref{fig:f1LL} shows the resulting phenomenological twist-2 distributions $f_{1LL}$ for valence and sea quarks.
The distributions are negative at small $x$, change sign at $x\simeq0.22$, and become positive at larger $x$.

\begin{figure}[tbp]
  \centering
  \begin{minipage}[t]{0.49\hsize}
    \includegraphics[clip,width=\linewidth]{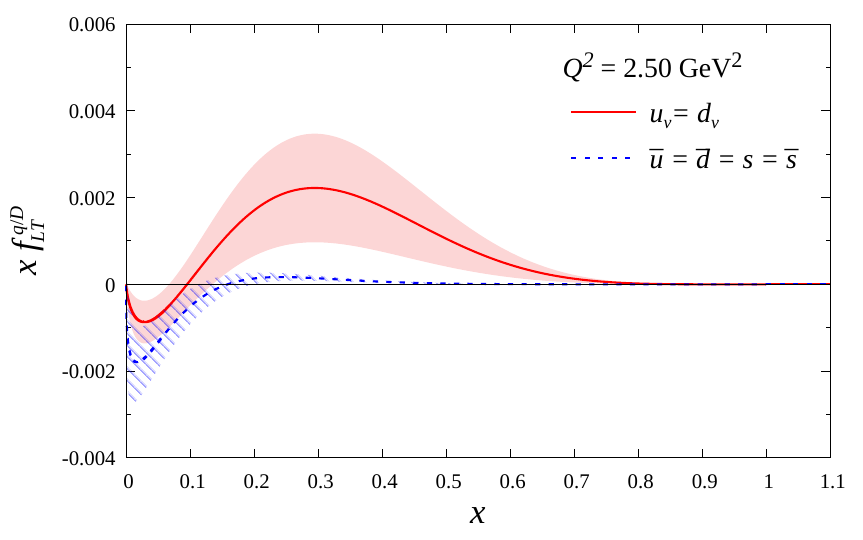}
  \end{minipage}
  \begin{minipage}[t]{0.49\hsize}
    \includegraphics[clip,width=\linewidth]{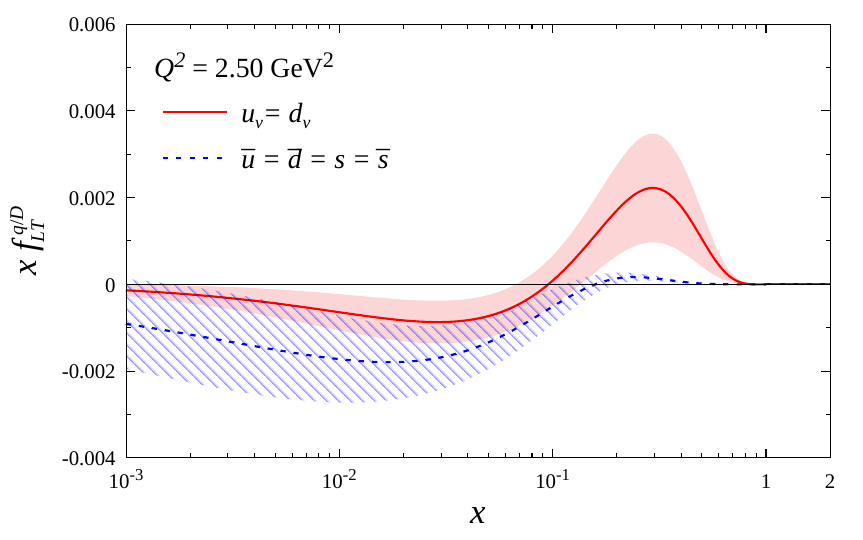}
  \end{minipage}
  \caption{
    Tensor-polarized twist-3 distributions $f_{LT}$ of the deuteron under the Wandzura-Wilczek approximation.
    The red solid lines represent the valence-quark distributions, while the blue dashed lines represent the sea-quark distributions.
    The left panel uses a linear scale for Bjorken-$x$, while the right panel uses a semi-logarithmic scale to better display the small-$x$ region.
  }\label{fig:fLT}
\end{figure}

Within the WW approximation, the twist-3 distribution $f_{LT}$ is given by
\begin{equation}
    f_{LT}(x) \simeq f_{LT}^\text{twist-2}(x) = \frac{3}{2} \int_{x}^2 \frac{\d y}{y} f_{1LL}(y).
\end{equation}
Using the phenomenological $f_{1LL}$ shown in Fig.~\ref{fig:f1LL}, we obtain the corresponding WW estimate of $f_{LT}$ shown in Fig.~\ref{fig:fLT}\@.
The twist-3 distribution $f_{LT}(x)$ has a shape similar to that of the twist-2 distribution $f_{1LL}(x)$.
In addition, the magnitude of $f_{LT}$ is of the same order as that of $f_{1LL}$\@.
Although subleading-twist contributions to cross sections are generally suppressed by $1/Q$ relative to leading-twist contributions, the sizable $f_{LT}$ suggests that tensor-polarized twist-3 effects may be accessible at relatively low $Q^2$, such as in experiments at JLab, where higher-twist effects can be more significant.

\section{Summary}

We have investigated the twist-2 and twist-3 tensor-polarized partonic structure of the deuteron, focusing on the relation between the leading- and subleading-twist distributions.
Using the operator product expansion with local operators, we derived a Wandzura-Wilczek-like twist-2 relation and a Burkhardt-Cottingham-like sum rule for tensor-polarized PDFs, taking advantage of the manifest Lorentz structure and rotational invariance of the local-operator formalism.
These relations provide useful theoretical constraints for investigating the twist-3 sector and clarify how the twist-3 distribution $f_{LT}$ is related to the leading-twist distribution $f_{1LL}$, while separating the genuine twist-3 contribution associated with multiparton correlations.
We then used a phenomenological parametrization of the twist-2 distribution constrained by the HERMES $b_1$ data to estimate the twist-3 distribution $f_{LT}$ of the deuteron within the Wandzura-Wilczek approximation.
The resulting WW estimate of $f_{LT}$ has a shape and magnitude comparable to those of $f_{1LL}$, suggesting that tensor-polarized twist-3 contributions may be relevant for future studies of subleading-twist spin structure.

\acknowledgments

SK and KK thank the Chinese Academy of Sciences for its support.
KK is also supported by the Gansu Province Postdoctoral Foundation.

\end{document}